\documentstyle[11pt]{article}

\begin{document}
\topmargin 0pt
\oddsidemargin 5mm
\setcounter{page}{1}
\begin{titlepage}
\hfill Preprint YERPHI-1537(11)-99

\vspace{2cm}
\begin{center}

{\bf Bound States  of the particles with Identical Charge in Magnetic Field }\\
\vspace{5mm}
{\large R.A. Alanakyan, L.A.Manukyan}\\
\vspace{5mm}
{\em Theoretical Physics Department,
Yerevan Physics Institute,
Alikhanian Brothers St.2,

 Yerevan 375036, Armenia\\}
 {E-mail: alanak@lx2.yerphi.am\\}
\end{center}

\vspace{5mm}
\centerline{{\bf{Abstract}}}

In this paper we consider  bound states and resonances of the particles with identical charge
in the  presence of the strong magnetic field.

\vspace{5mm}
\vfill
\centerline{{\bf{Yerevan Physics Institute}}}
\centerline{{\bf{Yerevan 1999}}}

\end{titlepage}

In this paper we consider  bound states and resonances of
particles with identical charge
 in the  presence of the strong magnetic field.
It is known that in the  presence of the strong magnetic field electron
 movement in Coulomb potential can be considered as one-dimensional.On the other
 hand, it is known that the term $-\alpha^2/(2mr^2)$ provide an attraction even for the
  positron.
Thus, in one-dimensional case this attraction leads to the existence of the bound states
of the same charge (e.g. $e^+p$,$lepton-lepton$,$antilepton-antilepton$ atoms),
because in one-dimensional case any attraction is enough for bound states formation.
Thus, we prove that in the strong magnetic field  bound states of the same charge exist.
Below we calculate energy levels of these bound states.

Dirac equation for an electron in case of attractive potential has been 
derived in [1][2] where, however, relativistic term
( the second term in the formula (2) below) has been neglected.In [3] this term
 has been taken into account.
However in [3] repulsive Coulomb potential has not been considered.
Analoguosly the Dirac equation for positron in the repulsive Coulomb 
potential with the presence of magnetic field ($e^+p$-atoms) has 
the following form:
\begin{equation}
\label{A1}
(\frac{-1}{2m}(\frac{d^2}{dr^2}+\frac{1}{r}\frac{d}{dr}+\frac{1}{r^2}\frac{d}{d\phi^2}
+\frac{1}{r}\frac{d^2}{dz^2}-\gamma^2r^2+2i\gamma\frac{d}{d\phi})+V(r,z))\psi(x)=E_{eff}\psi(x)
\end{equation}
where $E_{eff}=\frac{E^2-m^2}{2m}-\frac{eH}{2m}$,
\begin{equation}
\label{A3}
V(r,z)=\frac{E}{m}\frac{Z\alpha}
{\sqrt{r^2+z^2}}-\frac{Z^2\alpha^2}{2m}\frac{1}{r^2+z^2},
\end{equation}
$0<E<m$ Z-charge of nuclei.
Of course,in principle it is possible to solve this equation
numerically without any assumption.

In accordance with [1][2] if magnetic field is strong
(i.e. $\sqrt{\frac{1}{eH}}<<a_0=\frac{1}{m Z \alpha}$,where $a_0$ -
is Borh radius) transverse motion of electrons is defined only by
 magnetic field and we find the solution in  the following form:
\begin{equation}
\label{A4}
\psi(r,\phi,z)=\frac{1}{\sqrt{2\pi}}R_{00}(\rho)\chi(z)
\end{equation}
where $R_{00}(\rho)=\exp(-\frac{\rho}{2})$ is the function of the ground state $n=0,l=s=0$,
$\rho=\gamma r^2$ ($\gamma=\frac{eH}{2}$)whereas  $\chi(z)$ must be found below as a
solution to one-dimensional Schredinger equation.Substituting (2)(3) in (1)
 and multiplying by
$\frac{1}{\sqrt{2\pi}}R_{00}$ and integrating over $d^2r$ we obtain:
\begin{equation}
\label{A5}
(\frac{d^2}{2mdz^2}+(E_{eff}-\frac{\gamma}{2m})V(z))\psi(x)=0
\end{equation}
where
\begin{equation}
\label{A6}
V(z)=\frac{E}{m}\alpha\sqrt{\gamma}\int^{\infty}_0
\frac{d\rho\exp(-\rho)}{\sqrt{r^2+z^2}}-\frac{\alpha^2\gamma}{2m}\int^{\infty}_0\frac{d\rho\exp(-\rho)}{r^2+z^2}
\end{equation}
The behaviour of V(z) at different $\gamma$ is shown on Fig.1.
Thus, we have one-dimensional task with attraction at sufficiently
small $z$ which guarantees the existence of the bound states and
resonances.In [1][2] $\rho$ has been neglected in denominators in
formula for  $V(z)$ above. We however do not neglect it, because  at small $z$ it is not correct
and besides the presence of $\rho$ regularized the behaviour of the potential $V(z)$ at
small $z$.If we take into account formfactor of proton we obtain more
 smooth behaviour of potential at small $z$.

We solve this one dimensional Schredinger equation (4) numerically.
Our numerical results for energy of the ground state versus H at fixed  
charge of nuclei Z is shown on the Fig.2.

The author express his sincere gratitude to 
E.B.Prokhorenko for helpful discussions.

{\bf Figures Caption}

Fig.1 The behaviour of V(z) at different $\gamma$.(available after request)

Fig.2 Energy of the ground state versus $H$ at fixed $Z$(available after request)

\end{document}